\documentclass[cits]{PoS}

\title{Signals of the QCD Phase Transition in the Heavens}

\ShortTitle{Signals of the QCD Phase Transition in the Heavens}

\author{\speaker{J\"urgen Schaffner--Bielich}\\
Institut f\"ur Theoretische Physik/Astrophysik\\
J. W. Goethe Universit\"at\\
D-60438~Frankfurt am Main, Germany\\
E-mail: \email{schaffner@astro.uni-frankfurt.de}}

\abstract{The modern phase diagram of strongly interacting matter
  reveals a rich structure at high-densities due to phase transitions
  related to the chiral symmetry of quantum chromodynamics (QCD) and the
  phenomenon of color superconductivity. These exotic phases have
  a significant impact on high-density astrophysics, such as the
  properties of neutron stars, and the evolution of astrophysical systems
  as proto-neutron stars, core-collapse supernovae and neutron star
  mergers. Most recent pulsar mass measurements and constraints on
  neutron star radii are critically discussed. Astrophysical signals for
  exotic matter and phase transitions in high-density matter proposed
  recently in the literature are outlined. A strong first order phase
  transition leads to the emergence of a third family of compact stars
  besides white dwarfs and neutron stars. The different microphysics of
  quark matter results in an enhanced r-mode stability window for
  rotating compact stars compared to normal neutron stars. Future
  telescope and satellite data will be used to extract signals from phase
  transitions in dense matter in the heavens and will reveal properties
  of the phases of dense QCD. Spectral line profiles out of x-ray bursts
  will determine the mass-radius ratio of compact stars. Gravitational
  wave patterns from collapsing neutron stars or neutron star mergers
  will even be able to constrain the stiffness of the quark matter
  equation of state. Future astrophysical data can therefore provide a
  crucial cross-check to the exploration of the QCD phase diagram with
  the heavy-ion program of the CBM detector at the FAIR facility.}

\FullConference{
Critical Point and Onset of Deconfinement 4th International Workshop\\
July 9-13 2007\\
GSI Darmstadt,Germany}

\begin{document}


\section{Introduction}

\begin{figure}
\centering\includegraphics[width=0.7\textwidth]{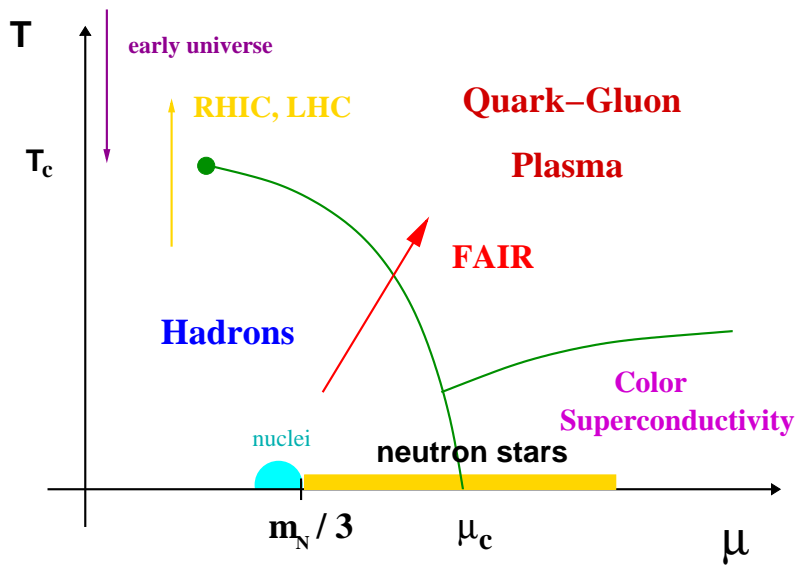}
\caption{A sketch of the phase diagram of QCD for different temperatures
  $T$ and quark chemical potential $\mu$. The heavy-ion program of FAIR
  at GSI Darmstadt will explore the phase transition line at high baryon
  densities.}
\label{Fig:phasediagram}
\end{figure}

The phase diagram of quantum chromodynamics (QCD) captures the bulk
properties of strongly interacting matter at extreme temperatures and
baryon number densities. A modern version of the QCD phase diagram is
sketched in Fig.~\ref{Fig:phasediagram}. There are striking similarities
to the phase diagram of water (just replace density with pressure) if
one associates the solid phase at small temperatures and high densities
with the colour-superconducting phase, the liquid phase at high
temperatures and densities with the quark-gluon plasma (QGP), and the
gas phase at low densities with the hadron gas. There is even a triple
point where all phases are in equilibrium and last but not least a
critical endpoint%
\footnote{Of course, the physics of the phase transitions is entirely
  different because the QGP is a plasma phase, not a liquid phase of
  hadrons.}.  
Note, that the QCD phase transition lines are related to the chiral
phase transition and symmetry arguments not to the deconfinement phase
transition. 

The phase diagram at small baryon density and high temperature will be
explored by the heavy-ion program of the LHC at CERN and is the physics
of the early universe at about $10^{-5}$~s after the big bang. The
heavy-ion program of CBM at the FAIR facility at GSI Darmstadt will
probe the properties of QCD matter at high baryon densities and
nonvanishing temperatures \cite{Friese2007}. Interestingly, this part of
the QCD phase diagram is of particular interest for high-density
astrophysics. Typical densities and temperatures encountered in
simulations of core-collapse supernovae, proto-neutron star evolution,
and neutron star mergers reach temperatures of 50~MeV and densities of
several times normal nuclear matter density, well in the range of the
estimates for the location of the phase transition line(s) of QCD. In
all these astrophysical systems, the nuclear equation of state (EoS)
serves as an essential ingredient, so that the QCD phase transition can
leave an imprint in astrophysical observations. Here, the existence of a
strong phase transition in a certain density and temperature range
suffices for our purposes not the exact location. An outline of possible
signals for the QCD phase transition in the heavens will be the topic of
this contribution thereby updating Ref.~\cite{SchaffnerBielich:2004ch}. The
study of the role of QCD in high-density astrophysics has emerged as an
extremely active field of research. On account of this, I apologise
beforehand, that the list of signals discussed here can neither be
exhaustive nor complete; a supplementary exposition with relations to
heavy-ion physics can be found in
\cite{Klahn:2006ir,Fuchs:2007vt,Sagert:2007nt,Blaschke:2007fr} and a
general review on strange quark matter in compact stars in
\cite{Weber:2004kj}.
 

\section{High-density astrophysics}


Matter under extreme densities can be found in astrophysical scenarios
involving compact objects. In core-collapse supernovae a star with a
zero-age main sequence mass of eight solar masses or more, collapses to a
hot proto-neutron star and then eventually to a black hole. The
modelling of core-collapse supernovae has evolved to new generations of
simulation codes, with the first three-dimensional simulations, modelling of
Boltzmann neutrino transport, and inclusion of rotation and magnetic
fields (for a recent review see e.g.~\cite{Janka:2006fh}). Earlier
models did not explode, so the question of missing physics was
raised \cite{Buras:2003sn}, the nuclear EoS was mentioned in this
context. Just recently, it turned out, that an instability can generate
a successful explosion, the standing accretion shock instability SASI
\cite{Janka:2007yu}. In any case, the possible role and impact of the
nuclear EoS for successful supernova explosions remains to be explored
in future simulations.

\begin{figure}
\centering\includegraphics[width=0.56\textwidth]{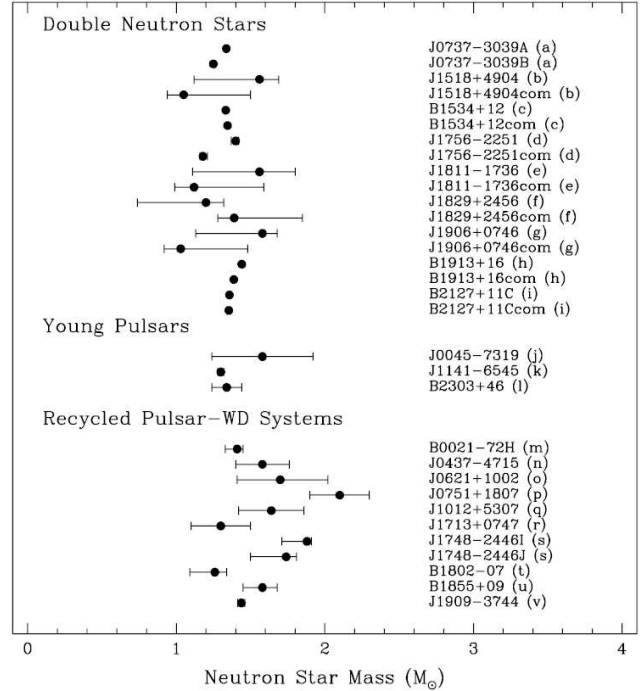}
\caption{Compilation of radio pulsar mass measurements (taken from
  \cite{Stairs:2006yr}, by courtesy of Ingrid Stairs).}
\label{Fig:pulsarmasses} 
\end{figure}

If the proto-neutron star cools down (and does not collapse to a black
hole) a neutron star is born in the aftermath of the core-collapse
supernova. These compact massive stars constitute the densest material
so far being directly observed in the sky. Typical radii of 10 km and
masses of one to two solar masses give {\em average} densities between
two and four times normal nuclear matter density. The density in the
core of a neutron star can be even substantially larger, which depends
solely on the behaviour of the high-density EoS. One key property of a
neutron star is its mass which can be measured quite precisely by
observing binary radio pulsars. A pulsar is a rotating neutron star, a
binary pulsar one with a companion star, which might be either a normal
star, a white-dwarf or another neutron star. If the companion star is a
white-dwarf or a neutron star, corrections from the Keplerian motion due
to effects from general relativity can be used to determine the mass
quite precisely, the accuracy is just a matter of observation time.

A rather recent compilation from 2006 of radio pulsars is depicted in
Fig.~\ref{Fig:pulsarmasses}, see \cite{Stairs:2006yr} for details.  More
than 1600 radio pulsars are known but only a few of them are binary
pulsars.  The mass of the famous Hulse-Taylor pulsar has now been
measured to $M=(1.4414\pm 0.0002)M_\odot$, the uncertainty being
comparable to the one for the gravitational constant $G$
\cite{Weisberg:2004hi}. Thanks to a recent flurry of activities in
detailed radio scans for pulsars, many new binary pulsars have been
discovered in the last few years and there are many more to come.
Particularly massive neutron stars might be in pulsar--white dwarf systems
due to accretion of the neutron star from its companion during the
evolutional history of the binary. Indeed, first measurements reported a
quite large mass for the pulsar J0751+1807 of $M=1.6-2.5M_\odot$
($2\sigma$) \cite{Nice:2005fi}. However, new data corrected the pulsar
mass to only $1.26M_\odot$ with a 68\% confidence range of 1.14 to 1.40
solar masses \cite{Stairs2007} (there was a problem with the on-line
folding ephemeris used for the oldest data which could only be detected
once new data were added). On the other hand, new mass limits were
announced at the Montreal conference on pulsars. The mass of the pulsar
J0621+1002 was determined to be $1.69M_\odot$ with a 68\% confidence
range of 1.53 to 1.80 solar masses \cite{Kasian2007}.  A new upper mass
limit has been extracted by combining data from the pulsars Terzan 5I
and J \cite{Ransom:2005ae} with the pulsar B1516+02B \cite{Freire2007}.
Under the assumption that the observed advances of periastron are due
only to general relativity (as opposed to companion mass quadrupole
moments), there is a 99\% probability that at least one of the three
neutron stars has a mass greater than $1.77 M_\odot$. Finally, the mass
measurement of the pulsar J1748--2021B with a median of $2.73M_\odot$
results in a lower mass limit of $M>2M_\odot$ with a 99\% confidence
level \cite{Freire2007}. If confirmed the latter result would be
sensational, but the total system mass is so large that it could in
principle also be a double neutron star system.

\begin{figure}
\centering\includegraphics[width=0.7\textwidth]{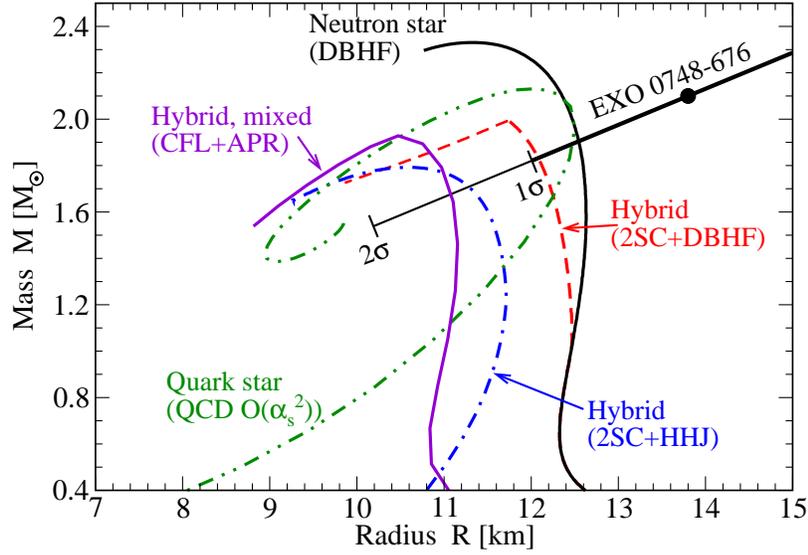}
\caption{Mass-radius relation for quark stars and hybrid stars showing
  that quark matter inside compact stars still allows for rather large
  masses and radii (taken from \cite{Alford:2006vz}).}
\label{Fig:MR_quark}
\end{figure}

Present limits on the radius and mass of neutron stars are more model
dependent (a very promising procedure of determining the mass and
redshift of a neutron star by relying just on effects of the warpage of
space-time will be given in the outlook on future observation
capabilities).  The recent analysis of the x-ray burster EXO 0748--676
\cite{Ozel:2006km} uses as input {\em inter alia} the redshift extracted
from the analysis of spectral lines \cite{Cottam:2002cu}. For a critical
assessment on the assumptions made in \cite{Ozel:2006km} I refer the
reader to Ref.~\cite{Walter2006}. In any case, the large mass and radius
extracted of $M\geq 2.10\pm 0.28 M_\odot$ and $R\geq 13.8 \pm 1.8$ km do
not rule out that unconfined quarks exist at the center of neutron stars
as claimed in \cite{Ozel:2006km}. In Ref.~\cite{Alford:2006vz}, modern
equations of state were used for pure quark stars as well as hybrid
stars, compact stars with an outer layer of neutron star matter and
quark matter in the core. Figure~\ref{Fig:MR_quark} shows the results
for the mass-radius relation together with the mass-radius constraint of
\cite{Ozel:2006km}. It clearly demonstrates that the mass-radius curves
pass through the mass-radius constraints given above and that quark
stars or hybrid stars can not be ruled out per se. However, a very soft
nuclear EoS would be incompatible with the data.

A rather common misconception relates the appearance of quark matter to
a substantial softening of the EoS which is actually not correct. The
EoS becomes softer for every new particle species appearing in nuclear
matter, as hyperons or kaon condensation, as a new degree of freedom
opens a new Pauli depth filling low-lying Fermi levels and therefore
lowering the total pressure for a given baryon density. The quark phase
on the other hand is a completely new phase, not a mere addition to the
hadronic degrees of freedom. With our present (poor) knowledge about the
properties of strongly interacting matter at high baryon densities, the
relation between pressure and energy density is basically unknown.
Interactions between quarks can result in rather large masses (see the
perturbative QCD calculation in Fig.~\ref{Fig:MR_quark}). That pure
quark stars can be quite massive is not a particularly new result, it
has been known at least since the classic works on strange stars within the MIT
bag model \cite{Witten:1984rs,Haensel:1986qb,Alcock:1986hz} with a
typical maximum mass of $2.1M_\odot$. However, there the stability is given by
the bag constant, i.e.\ the nontrivial vacuum of QCD. In either case,
the EoS of quark matter turns out to be strikingly similar in both
approaches \cite{Fraga:2001id} and large masses are possible for quark
stars (see below).


\section{Hunting down strange quark matter in the heavens}


There have been numerous signals proposed in the literature for the
presence of quark matter or a phase transition to it in astrophysical
systems. Some signals include an 'exotic' mass-radius relation of
compact stars, rapidly rotating pulsars due to an r-mode {\em stability}
window, enhanced cooling of neutron stars (see the contribution of David
Blaschke \cite{Blaschke2007}), and gamma-ray bursts by transition to
strange quark matter (GRBs without a supernova, late x-ray emission,
long quiescent times etc.).  For gravitational wave signals of phase
transitions, astrophysical systems of interest involve binary neutron
star collisions, collapse of a neutron star to a hybrid star, and the
r-mode spin-down of hybrid stars.

In fact, the notion of the QCD phase transition has been modelled in the
past mainly using the MIT bag model or the assumption that there is a
transition from hadronic matter to (free) quark matter. Modern
investigations are more based on the underlying properties of dense QCD,
incorporating that the true transition in dense QCD originates from chiral
symmetry rather than deconfinement. The phase diagram of QCD for
astrophysical applications has been worked out in detail in
\cite{Ruster:2005jc} incorporating effects from color-superconductivity
in $\beta$ equilibrium and in neutrino-trapped matter
\cite{Ruster:2005ib}.  Besides the phases and phase transition lines
depicted in Fig.~\ref{Fig:phasediagram} many more have been found, due
to different pairing patterns of the quarks: normal quark phase,
two-flavor color superconducting phase (2SC), gapless 2SC phase,
color-flavor locked phase (CFL), gapless CFL phase, and metallic CFL
phase (for the original literature see
\cite{Alford:1997zt,Rapp:1997zu,Alford:1998mk,Shovkovy:2003uu,Alford:2003fq}).

It should be stressed again that the existence of the phases and phase
transition lines are based on symmetry arguments! However, the location
in the phase diagram can not be fixed in a model independent way, so that it
is not clear which phase is realized in nature e.g.\ in the core of a
neutron star or a proto-neutron star. Unfortunately, this statement
implies that the EoS for quark matter is equally undetermined as well
as its matching to the low density hadronic part. Some basic insights
can be extracted by considering pure quark matter and pure quark stars
first.

\begin{figure}
\centering\includegraphics[width=0.6\textwidth]{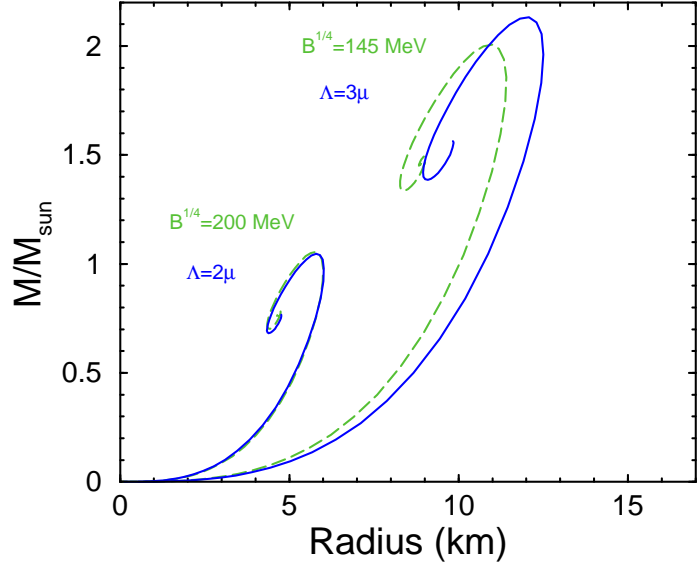}
\caption{The mass-radius relation for pure quark stars within the MIT
  bag model (long-dashed lines) and within perturbative QCD calculation
  to order $\alpha_s^2$ (solid lines, see
  \cite{Fraga:2001id,Fraga:2001xc} for details).}
\label{Fig:mr_quarkbag}
\end{figure}

The typical mass-radius relation for pure quark stars (strange stars) is
shown in Fig.~\ref{Fig:mr_quarkbag} for the MIT bag model (see
\cite{Haensel:1986qb,Alcock:1986hz}) and for a calculation using
perturbative QCD up to order $\alpha_s^2$ \cite{Fraga:2001id}. The
compact star mass increases with the radius $M\propto R^3$, i.e.\ the
average density is about constant. The maximum mass depends crucially on
the chosen value of the MIT bag constant or the choice of the
renormalisation subtraction point $\Lambda$. Interestingly, the
mass-radius curves of both approaches lie very close to each other,
despite the completely different approaches. Note, that the maximum mass
of quark stars can reach values of $2.1 M_\odot$ with a radius of 12 km,
the same typical values as for ordinary neutron stars.  The mass-radius
relation of pure quark stars is actually quite generic, as they are
selfbound stars stabilised by interactions not by gravity.  The common
features are that the pressure vanishes at a finite energy density, so
that the mass-radius relation starts at the origin (ignoring a possible
crust) and arbitrarily small masses and radii are possible. The
mass-radius curves follow a scaling relation proportional to the
critical energy density where the pressure vanishes for the MIT bag
model the mass and radius scale with the bag constant as $B^{1/2}$
\cite{Witten:1984rs}. In contrast, ordinary neutron stars are bound by
gravity, the pressure is nonvanishing for a nonvanishing energy density,
so that the mass-radius relation starts at large radii with a minimum
neutron star mass of $M\sim 0.1M_\odot$ at $R\sim 200$ km.  Strange
stars have similar masses and radii, and if there is a nuclear crust,
also similar surface properties. However, there are astrophysical
observables which could reveal the existence of a strange star or
strange quark matter, as extremely small compact star masses and radii,
small and light white dwarfs (strange dwarfs) with unusual mass-radius
relation due to a strange star core
\cite{Glendenning:1994sp,Mathews:2006nq}, super-Eddington luminosity
from bare, hot strange stars \cite{Page:2002bj} and explosive events due
to the conversion of neutron stars to strange stars. However, pure quark
stars can only exist if the strange matter hypothesis holds, which
hypothesises that strange quark matter is absolutely stable.

For a probably more realistic scenario one has to match quark matter to
hadronic matter, so that the compact star consists of both types of
matter, a so-called hybrid star. To be more precise here, one has to
take into account the transition from the chirally restored phase
('quark matter') to the chirally broken one ('hadronic matter'). For
simplicity, I stick to the traditional terms of quark matter and
ordinary hadronic matter in the following. If the chiral phase
transition is of first order as suggested by the QCD phase diagram,
Fig.~\ref{Fig:phasediagram}, there are two extreme possibilities for the
matching. Either, the chiral transition is weakly first order (or there
is no true phase transition), then there is just one type of compact
star (an ordinary neutron star). Or the chiral phase transition is
strongly first-order, then two types of compact stars are possible with
a new stable solution at smaller radii. This new stable solution is
sometimes called a quark star twin or a third family of compact stars
\cite{Gerlach68,Kaempfer81,Haensel82,Glendenning:1998ag,Schertler:2000xq}.
Note, that stars of the third family are hybrid stars. The two different stable
solutions can be distinguished by their core composition, the second
family has just a mixed phase in the core while the third family
consists of a pure quark matter core. Note, that the new solution to the
Tolman-Oppenheimer-Volkoff equations is not only stable but is possible
for any kind of first order phase transition (also hypothetical hadronic
ones, see \cite{SchaffnerBielich:2002ki}). That means also that a strong
phase transition within color-superconducting quark matter can
generate compact star twins, like the transition from the 2SC to the CFL
phase \cite{Sandin:2007zr}.

There have been several astrophysical signals proposed specifically for
a phase transition and/or a third family of compact stars as e.g.: the
spontaneous spin-up of pulsars \cite{Glendenning:1997fy,Zdunik:2005kh},
the 'rising twin' feature of the mass-radius relation
\cite{Schertler:2000xq}, the existence of sub-millisecond pulsars and
gravitational wave bursts due to r-mode (in)stabilities
\cite{Drago:2004nx}, and gamma-ray bursts with late x-ray emission and
long quiescent times \cite{Drago:2005rc}. Other possible observables
include the emission of gravitational waves, $\gamma$-rays, and neutrino
from the collapse of a neutron star to the third family, a possible
secondary shock wave in supernova explosions, as well as gravitational
waves from colliding neutron stars.

The observation of a 1122Hz signal from the low-mass x-ray binary XTE
J1739 \cite{Kaaret:2006gr} triggered a reanalysis of the limiting
rotation frequencies of compact stars (note that it is not clear at
present whether this frequency really corresponds to the rotation
frequency of the pulsar, unfortunately). Besides the standard Kepler
limit, the compact star can only rotate fast in a certain temperature
range (see \cite{Madsen:1999ci} and references therein). Other regions
are excluded by the so-called r-mode instability, i.e.\ the compact star
slows down rapidly due to the emission of gravitational waves. This
instability window depends on the microphysics of dense matter,
hence, it is different for ordinary nuclear matter and for quark matter.
Accreting neutron stars have typical inner temperatures of $T=10^8$ K or
higher.  Ordinary neutron star matter (even with hyperons) would be
unstable with respect to the r-mode at a high rotation frequency around
this temperature. On the other hand, hybrid stars are stable thanks to
the significantly different bulk viscosity of quark matter (see e.g\
\cite{Sa'd:2007ud}). Indeed, it was demonstrated by \cite{Drago:2007iy}
that only compact stars with quark matter can rotate that fast at a
temperature of $T=10^8$~K . These are exciting prospects and the
confirmation of the existence of an accreting submillisecond pulsar
is eagerly awaited.

Colliding neutron stars, stars collapsing to black holes or collapsing
compact stars are all candidates for the sources of gamma-ray bursts.
These highly energetic events in the sky release energies similar to the
ones of supernovae and occur about once per day. There is a special
subclass of gamma-ray burst with long quiescent times (more than 40
seconds) between two bursts \cite{Drago:2005rc}. The characteristics of
the two bursts before and after quiescence are quite similar, which
points towards a dormant inner engine and not to different physics as
would be the case for shocks outside the inner engine. A possible source
of the gamma-ray bursts could then be a compact star experiencing two
consecutive phase transitions: first to quark matter (chiral phase
transition) and then to color-superconducting matter. The energy release
of both QCD transitions is huge and well in the range of that extracted
from gamma-ray burst spectra.


\section{The hunt for astrophysical signals of QCD 
in the near future}  


There are presently a large number of telescopes and satellites relevant
for the observation of the properties of QCD in the heavens, such as the
radio telescopes at the Arecibo, Parkes, Jodrell Bank, and Green Bank
Observatories,
the Hubble Space Telescope, the Very Large Telescope (VLT) of the
European Southern Observatory, and the Keck Telescopes at Hawaii in the
optical,  
the x-ray satellites Chandra and XMM-Newton, 
as well as the gravitational wave
observatories GEO600, LIGO, and VIRGO, and 
the neutrino telescopes of
Super-Kamiokande and the Sudbury Neutrino Observatory (SNO).
The sensitivities in all these areas will be boosted and extended by
future telescopes, observatories and satellites: for radio observation
the Square Kilometre Array (SKA), for optical observations the James
Webb Space Telescope (JWST), the successor of the Hubble Space
Telescope, and the Extremely Large Telescope, for x-ray observations the
satellite missions Constellation-X and XEUS, the Laser Interferometer
Space Antenna (LISA) for the detection of gravitational waves, and the
Underground Nucleon decay and neutrino Observatory (UNO) for the
detection of even extragalactic supernova neutrinos.

A few astrophysical observables proposed which can be measured with
those future detectors are listed in the following.

For the first time, a double pulsar system (PSR J0737-303) was detected,
which is a system of two neutron stars where the radio signal of both
pulsars has been measured \cite{Lyne:2004cj}. This system at present
provides the best tests of General Relativity in the strong field regime
\cite{Kramer:2006nb}. For double pulsar systems, the moment of inertia
can be measured, which will give constraints on the radius of neutron
stars \cite{Morrison:2004df,Lattimer:2004nj,Bejger:2005jy}. The mass of
the pulsar of interest is already very precisely known $M_A=1.3381(7)
M_\odot$ \cite{Kramer:2006nb}. If the moment of inertia can be
determined with an accuracy of 5 to 10\% it will provide a tight
constraint on the mass-radius relation and on the nuclear EoS just by
using general relativity
\cite{Morrison:2004df,Lattimer:2004nj,Bejger:2005jy}.

Accreting neutron stars can reveal themselves by x-ray bursts which
originate from the surface of the neutron star. The profile of the
emitted spectral lines are modified by the strong space-time warpage
which depends on the compactness of the compact star, the ratio of the
mass to the radius. The warped spectral shape can then be used as a
model-independent measurement of the mass-radius relation of compact
stars. With Constellation-X the mass-radius ratio is expected to be
narrowed down to within 5\% accuracy
\cite{Strohmayer:2004jf,Bhattacharyya:2005ge}. The method has been
applied successfully for the pulsar J0437--4715 by using the present
x-ray data. The radius is limited within 99.9\% confidence to $R>6.7$~km
for an assumed mass of $1.4 M_{\odot}$ \cite{Bogdanov:2006zd}.

Gravitational wave astronomy has opened up an entirely new window to
'listen' to the universe. Sources of gravitational waves are suspected
to be, for example: nonspherical rotating neutron stars, colliding
neutron stars and black-holes, and supernovae. Gravitational wave
detectors are observing right now (as LIGO and GEO600). They just set
new limits on the gravitational wave emission of pulsars getting close
to the spin-down limit \cite{Abbott:2007ce}. Quark matter in the core of
neutron stars could in fact sustain large deformations and would be
visible in a sizable emission of gravitational waves
\cite{Owen:2005fn,Haskell:2007sh}. The spin-down of hybrid stars due to
r-mode instabilities could lead to an observable gravitational wave
burst \cite{Drago:2004nx}. There are several numerical simulations of
gravitational wave emission reported in the literature which include
effects from a phase transition to quark matter at high densities.  In
binary neutron star mergers with a quark core, a signal is clearly seen
in the Fourier spectrum of the gravitational wave signal
\cite{Oechslin:2004yj}.  In modelling binary strange quark star
collision, higher frequencies are possible before 'touch-down' in
comparison to normal neutron stars \cite{Limousin:2004vc}. The collapse
of an ordinary neutron star to a hybrid star with a quark matter core
produces characteristic gravitational waves
\cite{Lin:2005zd,Yasutake:2007st}. The authors adopt a simple polytropic
EoS for the quark matter phase with an adiabatic index of $\Gamma=1.75$
to 1.95.  Remarkably, the pattern of the gravitational wave signal turns
out to be highly sensitive to the stiffness of the quark matter EoS
\cite{Lin:2005zd,Yasutake:2007st}. Finally, the collapse of neutron
stars to hybrid stars due to a phase transition in neutron stars will
generate a gravitational wave background \cite{Sigl:2006ur}. The
background of such gravitational waves could be detectable with future
space based detectors. The signal could be even larger than the one for
conventional type II supernovae \cite{Sigl:2006ur}.


\section{Summary}


There is a unique one-to-one relation between the nuclear equation of
state and the mass-radius relation of compact stars.  The various phase
transitions in dense QCD matter lead to a rich variety of astrophysical
signals to be revealed by e.g.\ exotic mass-radius relation (the
existence of a third family of compact stars), x-ray observations of
x-ray binaries, gamma-ray bursts, gravitational wave emissions and
observations of core collapse supernovae.  There are lots of
opportunities for a cross-check between heavy-ion physics and
high-density astrophysics. Clearly, the future is bright for probing the
densest material known in the universe in the coming years with the CBM
detector on earth and with astrophysical detectors in the heavens.


\acknowledgments

I thank my colleagues and friends for many fruitful discussions on the
physics of QCD matter in the last few years and during this conference, in
particular I thank Alessandro Drago for providing me with a long list of
possible signals for quark matter in the sky. I am indebted to Ingrid
Stairs for a detailed update on the most recent and unpublished pulsar mass
measurements and clarifying corrections to the manuscript. Finally, I
thank Matthias Hempel, Giuseppe Pagliara, and Irina Sagert from the
compact star group in Frankfurt for numerous conversations and for their
collaboration.


\bibliographystyle{utcaps}
\bibliography{all,literat,cpod2007}

\end{document}